\documentclass[%
twocolumn,
superscriptaddress,
amsmath,amssymb,
prl,
raggedbottom
]{revtex4}

\usepackage{graphicx}
\usepackage{dcolumn}
\usepackage{bm}
\usepackage{comment}
\usepackage{amssymb}
\usepackage{amsmath}
\usepackage{amsthm}
\usepackage{mathtools}
\usepackage[usenames,dvipsnames]{xcolor}
\usepackage{epsfig}
\usepackage{dcolumn}
\usepackage{tikz}
\usepackage{braket}
\usepackage[export]{adjustbox}
\usepackage{hyperref}
\usepackage{verbatim}
\hypersetup{
	colorlinks=true,
	urlcolor=RoyalBlue,
	linkcolor=orange!80!black,
	citecolor=RoyalBlue,
	bookmarks=true,
	pdftitle={Landau Singularities Revisited},
	pdfauthor={Claudia Fevola, Sebastian Mizera, Simon Telen},
	pdfdisplaydoctitle=true,
	pdfstartview=FitH
}
\usepackage{orcidlink}
\usepackage[cachedir={_minted-JHEP},finalizecache=false,frozencache=true]{minted}
\setminted[julia]{frame=lines,
rulecolor=\color{white!80!black},
fontsize=\small,
numbers=right,
numbersep=-5pt,
obeytabs=true,
tabsize=4,
escapeinside=??}

\usetikzlibrary{snakes}
\usetikzlibrary{calc}
\usetikzlibrary{decorations.pathmorphing}

\tikzset{
    quark/.style={
        decoration={},
        decorate
    },
    lepton/.style={
        decoration={},
        decorate
    },
    gluon/.style={
        decoration={coil, aspect=0.75, mirror, segment length=1.5mm},
        decorate
    },
    gluon_small/.style={
        decoration={coil, aspect=0.75, mirror, segment length=1mm, amplitude =0.6mm},
        decorate
    },
    vector/.style={
        decoration={snake, aspect=0.75, mirror, segment length=2mm},
        decorate
    },
    Higgs/.style={
        decoration={},
        densely dashed,
        decorate
    },
}

\definecolor{color2}{rgb}{0.368417, 0.506779, 0.709798}
\definecolor{color3}{rgb}{0.880722, 0.611041, 0.142051}
\definecolor{color5}{rgb}{0.560181, 0.691569, 0.194885}
\definecolor{color1}{rgb}{0.922526, 0.385626, 0.209179}
\definecolor{color6}{rgb}{0.528488, 0.470624, 0.701351}
\definecolor{color4}{rgb}{0.772079, 0.431554, 0.102387}

\def \be {\begin{equation}}
\def \ee {\end{equation}}
\def \U {\mathcal{U}}
\def \F {\mathcal{F}}
\def \I {\mathcal{I}}

\def \L {\mathrm{L}}

\def \D {\mathrm{D}}
\def \d {\mathrm{d}}
\def \leq {\leqslant}
\def \geq {\geqslant}
\def \Newt {\mathrm{Newt}}
\def \nn {\nonumber}

\def \G {\mathcal{G}}

\begin{document}

\title{Landau Singularities Revisited:\\Computational Algebraic Geometry for Feynman Integrals}

\author{Claudia Fevola \orcidlink{0000-0001-5050-9209}}%
\affiliation{%
Université Paris-Saclay, Inria, 91120 Palaiseau, France
}

\author{Sebastian Mizera \orcidlink{0000-0002-8066-5891}}%
\affiliation{%
Institute for Advanced Study, Einstein Drive, Princeton, NJ 08540, USA
}

\author{Simon Telen \orcidlink{0000-0002-3459-5845}}%
\affiliation{%
Max Planck Institute for Mathematics in the Sciences, Inselstra\ss e 22, 04103 Leipzig, Germany
}

\begin{abstract}
We reformulate the analysis of singularities of Feynman integrals in a way that can be practically applied to perturbative computations in the Standard Model in dimensional regularization. After highlighting issues in the textbook treatment of Landau singularities, we develop an algorithm for classifying and computing them using techniques from computational algebraic geometry. We introduce an algebraic variety called the principal Landau determinant, which captures the singularities even in the presence of massless particles or UV/IR divergences. We illustrate this for $114$ example diagrams, including a cutting-edge $2$-loop $5$-point non-planar QCD process with multiple mass scales. The algorithms introduced in this work are implemented in the open-source \texttt{Julia} package \texttt{PLD.jl} available at \url{https://mathrepo.mis.mpg.de/PLD/}. 
\end{abstract}

\maketitle
\section{Introduction}

Modern high-precision computations in particle physics rely on the knowledge of the singularity structure of perturbative scattering amplitudes. This emphasizes the importance of being able to predict it without the explicit evaluation of Feynman integrals. A pioneering step in this direction was taken by Bjorken, Landau, and Nakanishi \cite{Bjorken:1959fd,Landau:1959fi,10.1143/PTP.22.128}, who demonstrated that poles and branch points of finite Feynman integrals can be determined by pinches of on-shell hypersurfaces in the loop-momentum space. They are known as \emph{Landau singularities}; see \cite[Sec.~1]{PrincipalLandauDeterminants} for a historical overview and more complete references. Nevertheless, it is not known how to \emph{systematically} apply this analysis to Standard Model processes beyond one-loop level, due to many problems outlined below. The goal of this letter is to fix these issues and give a \emph{practical} algorithm for determining singularities (physical or not) in scenarios with massless particles and UV/IR divergences.

Landau analysis is particularly important in the context of finding analytic expressions, differential equations, and symbol calculus for Feynman integrals, see, e.g., \cite{Badger:2023eqz}. As a simple example, if the amplitude has a term of the form $\frac{1}{P}\log [ (Q + \sqrt{R})/(Q - \sqrt{R})]$ for some polynomials $P,Q,R$ in the external kinematics, then its Landau singularities are located at $P=0$ (simple pole), $R=0$ (square root), and $Q^2 - R = 0$ (logarithmic) on some Riemann sheets. By definition, Landau singularities coincide with zeros and singularities of symbol letters in polylogarithmic cases \cite{Goncharov:2010jf,Maldacena:2015iua}. By basic elimination theory, their positions can be always expressed as $\Delta = 0$, where $\Delta$ is a polynomial in the Mandelstam invariants and masses-squared with coefficients in $\mathbb{Q}$. The challenge is to determine all such $\Delta$'s without having to compute Feynman integrals.

According to textbooks, positions of singularities of Feynman integrals can be computed by solving \emph{Landau equations} \cite{Eden:1966dnq,bjorken1965relativistic,nakanishi1971graph,Itzykson:1980rh,Sterman:1993hfp}. The \emph{leading} singularity is computed by setting all propagators on-shell and imposing that all interaction vertices are located at space-time points. \emph{Subleading} singularities are leading singularities of reduced diagrams obtained by contracting any subset of edges. In addition, \emph{second-type} and \emph{mixed} singularities are those corresponding to one or more loop momenta diverging. All of these are often summarized as solving the equations
\begin{subequations}\label{eq:textbook-LE}
\begin{align}
\alpha_i (q_i^2 - m_i^2) = 0 \quad&\text{for all internal edges}\; i\, ,\label{eq:textbook-LE-a} \\
\sum_{i \in a} \pm \alpha_i q_i^\mu = 0 \quad&\text{for all loops}\; a\, ,\label{eq:textbook-LE-b}
\end{align}
\end{subequations}
where $q_i^\mu$, $m_i$, and $\alpha_i$ are respectively the momentum, mass, and the Schwinger parameter of the $i$-th propagator, and the sum goes over all edges belonging to the $a$-th loop with signs denoting orientations.

Our work starts with an observation that the above classification is not complete. Indeed, there are known examples where a naive application of \eqref{eq:textbook-LE} does not detect all singularities and more careful blow-ups are needed \cite{Landshoff1966,10.1063/1.1724262,Berghoff:2022mqu}.
There are two fundamental problems with the above textbook lore:
(i) it is not clear why only $\alpha_i \neq 0$ or $\alpha_i = 0$ solutions are considered, in contrast with more complicated scaling patterns; and (ii) the prescription is ambiguous in the presence of UV/IR divergences, i.e., when \eqref{eq:textbook-LE} has solutions for any external kinematics.
In this work, building on \cite{Mizera:2021icv}, we introduce tools from \emph{nonlinear algebra} \cite{michalek2021invitation}, which tackle these issues as follows. 

\paragraph*{\bf (i) Beyond the standard classification.}
We find that, in general, a singularity can come from scalings
\be\label{eq:scaling}
\alpha_i \to \varepsilon^{w_i} \alpha_i \qquad\mathrm{with}\qquad \varepsilon \to 0\, .
\ee
Each region is given by the set of rational exponents $(w_1, w_2, \ldots)$ called \emph{weights}. The internal momenta $q_i^\mu$ will also have a specific scaling as a consequence of \eqref{eq:scaling}, which will be given later in \eqref{eq:scaling-momenta}. Their combinatorics is captured by a lattice polytope $\mathbf{P}$ described below. 

Hence, it is not enough to study reduced diagrams, but instead a whole suite of configurations obtained by shrinking and expanding edges at specific relative rates is needed. Below, we will explain that scalings beyond \eqref{eq:scaling} are often possible as well. This fact may not be surprising to the readers familiar with the method of regions \cite{Jantzen:2011nz}, which also exploits similar scalings, though only for $\alpha_i \geq 0$, while we consider any $\alpha_i \in \mathbb{C}$.

\paragraph*{\bf (ii) Interplay with UV/IR divergences.} The second issue is best illustrated on an example of a reduced diagram obtained by contracting a number of edges, say in a given QCD process with all external legs on-shell:
\be\label{eq:contraction}
\begin{gathered}
\begin{tikzpicture}[line width=0.7,scale=0.9]
	\begin{scope}
	\coordinate (v1) at -(0.2,-0.2);
	\coordinate (v7) at -(-0.3,0.5);
	\coordinate (v2) at -(0.2,1.2);
	\coordinate (v3) at -(1,1);
	\coordinate (v4) at -(2,1);
	\coordinate (v5) at -(2,0);
	\coordinate (v6) at -(1,0);
	\draw[gluon_small] (v6) -- (v1) -- (v7) -- (v2) -- (v3);
	\draw[quark] (v3) -- (v4) -- (v6);
	\filldraw[white] (1.5,0.5) circle (4pt);
	\draw[quark] (v3) -- (v5);
	\draw[quark] (v5) -- (v6);
	\draw[gluon_small] (v1) -- ++(-135:0.5);
	\draw[gluon_small] (v2) -- ++(135:0.5);
	\draw[Higgs] (v4) -- ++(45:0.6);
	\draw[gluon_small] (v5) -- ++(-45:0.5);
	\draw[gluon_small] (v7) -- ++(-180:0.5);
	\foreach \point in {v1, v2, v3, v4, v5, v6, v7} {
		\fill[white!20!black] (\point) circle [radius=0.07];
	}
	\draw[draw=orange!80!black, fill=orange!80!black, fill opacity=0.1] plot [smooth cycle, tension=0.5] coordinates {(-0.5,0.7) (1,0.5) (2.1,1.2) (2.1,-0.1) (0.1,-0.3)};
	\end{scope}
\begin{scope}[xshift=55mm]
	
	\draw[->] (-2.5,0.5) -- node[above,color=orange!80!black] {\footnotesize{contract}} (-1,0.5);
	
	\coordinate (v2) at (0,1);
	\coordinate (v3) at (1,1);
	\coordinate (v6) at (1,0);
	
	\draw[gluon_small] (v2) -- ++(180-25:0.9);
	\draw[Higgs] (v6) -- ++(25:0.75);
	\draw[gluon_small] (v6) -- ++(-110:0.8);
	\draw[gluon_small] (v6) -- ++(-25:0.75);
	\draw[gluon_small] (v6) -- ++(-155:0.9);
	
	\draw[gluon_small] (v2) -- (v3);
	\draw[gluon_small] (v2) to[out=-90,in=-180] (v6);
	\draw[quark] (v3) to[out=-60,in=60] (v6);
	\draw[quark] (v3) to[out=-120,in=120] (v6);
	
	\fill[white!20!black] (v2) circle [radius=0.07];
	\fill[white!20!black] (v3) circle [radius=0.07];
	\draw[fill=orange!80!black, draw=orange!80!black] (v6) circle [radius=0.09];
\end{scope}
\end{tikzpicture}
\end{gathered}
\ee
In this case, the diagram on the right-hand side is always singular due to a collinear divergence between gluons (curly) and possibly quarks (solid) if they are taken to be massless. Hence, the Landau equations in the form \eqref{eq:textbook-LE} have solutions regardless of the remaining external kinematics, i.e., they evaluate to $0 = 0$. However, it does \emph{not} mean we can simply discard them, because the same region of phase-space will in general lead to other genuine kinematic singularities. Moreover, it is clear one could not find them solely by considering the reduced diagram in \eqref{eq:contraction} which does not depend on any Mandelstam invariants. Instead, the limit \eqref{eq:contraction} needs to be taken more carefully.

More generally, we explain how to disentangle the divergences that are always present (UV/IR) from singularities that exist only for special values of the kinematics (Landau) using a combination of elimination theory and numerical irreducible decomposition.

The goal of this letter is to summarize the main construction in a way accessible to physicists and introduce the software. The follow-up paper \cite{PrincipalLandauDeterminants} will provide mathematical foundations and algorithmic details.

\section{Landau equations revisited}

A family of Feynman integrals with $n$ external legs in $\D$ dimensions can be parametrized by
\be\label{eq:family}
\I(z) := \int \frac{\d^\D \ell_1\, \d^\D \ell_2 \cdots \d^\D \ell_\L}{P_1^{\nu_1} P_2^{\nu_2} \cdots P_m^{\nu_m}}\, ,
\ee
where the $P_i$'s correspond to $m$ denominators and irreducible scalar products, each raised to a (possibly non-integer) power $\nu_i$, and $\ell_a$'s are the $\L$ loop momenta. We collectively call all the kinematic parameters (Mandelstam invariants and masses-squared) $z \in \mathcal{E}$, where $\mathcal{E}$ is the specific subspace of the kinematics we consider. The overall normalization does not affect the singularity structure. Introducing $m$ Schwinger parameters $\alpha = (\alpha_1, \alpha_2, \ldots, \alpha_m)$, we write
\be\label{eq:sum-alpha}
\sum_{i=1}^{m} \alpha_i P_i = \sum_{a,b=1}^{\L} \ell_a \cdot \ell_b\, \mathbf{Q}_{ab} + 2 \sum_{a=1}^{\L} \ell_a \cdot \mathbf{L}_a + c\, ,
\ee
where $\mathbf{L}$ and $c$ can depend on the internal masses and external kinematics, but are independent of the loop momenta.
The matrix $\mathbf{Q}$ and vector $\mathbf{L}$ give rise to the Symanzik polynomials:
\be
\U(\alpha) := \det \mathbf{Q}, \qquad \F(\alpha; z) := \left( \mathbf{L}^\intercal \mathbf{Q} \mathbf{L} - c \right) \U\, .
\ee
Hence, $\F$ depends on external kinematic invariants and masses, while $\U$ does not.
Following standard steps, the integrals in \eqref{eq:family} are then proportional to \cite{Lee:2013hzt}
\be\label{eq:UF-integral}
\I \propto \int_{\mathbb{R}^m_{+}} \!\,\frac{\d^m \alpha}{\G^{\D/2}}\, \alpha_1^{\nu_1 - 1} \alpha_2^{\nu_2 - 1} \cdots \alpha_m^{\nu_m - 1} \, ,
\ee
where
\be
\G(\alpha; z) := \U + \F / \mu^2
\ee
is often called the \emph{graph polynomial}. We fix the auxiliary mass scale $\mu=1$ from now on.
Singularities can also be analyzed in other representations, which leads to equivalent equations \cite[Sec.~3.4]{PrincipalLandauDeterminants}. One can show that propagators with $\nu_a \leq 0$ do not give rise to new singularities and can be excluded from the analysis \cite[App.~B]{PrincipalLandauDeterminants}.

The simplest Landau singularity (known as leading second-type) is obtained by solving the critical point equations away from the boundaries:
\begin{equation}
\label{eq:LE}
\G = \partial_{\alpha} \G = 0\quad \text{for}\quad \alpha \in (\mathbb{C}^\ast)^m \;\text{and}\; z \in \mathcal{E}\, ,
\end{equation}
where $\partial_\alpha = (\tfrac{\partial}{\partial \alpha_1}, \tfrac{\partial}{\partial \alpha_2}, \ldots, \tfrac{\partial}{\partial \alpha_m})$ and $\mathbb{C}^\ast = \mathbb{C} \setminus \{0\}$. It is the condition for the denominator in \eqref{eq:UF-integral} to be singular in a way that cannot be cured by contour deformations. The Feynman integral $\I(z)$ possibly develops a singularity whenever a solution to \eqref{eq:LE} exists for a given value of $z$. Back in the loop-momentum space, completing the square in \eqref{eq:sum-alpha} is equivalent to solving the loop Landau equations \eqref{eq:textbook-LE-b}. Hence, one can read off the behavior of the loop momenta
\be\label{eq:scaling-momenta}
\ell_a = -\sum_{b=1}^{\L} [\mathbf{Q}^{-1}]_{ab} \mathbf{L}_b
\ee
in terms of Schwinger parameters for every solution. Similarly, solving \eqref{eq:LE} imposes the on-shell conditions \eqref{eq:textbook-LE-a}.

Most of the singularities are hiding on the boundaries.
They can be classified using the Newton polytope:
\be
\mathbf{P} = \Newt\big(\G\big)\, .
\ee
The same polytope appears in other recent \emph{tropical} approaches to Feynman integrals \cite{Borinsky:2020rqs,Heinrich:2021dbf,Arkani-Hamed:2022cqe,Gardi:2022khw,Borinsky:2023jdv}. Its faces can be parametrized by weights $w = (w_1,w_2,\ldots, w_m)$ and are in one-to-one correspondence with scalings $\alpha_i \to \varepsilon^{w_i} \alpha_i$ with $\varepsilon \to 0$. In particular, the behavior of the loop momenta $\ell_a$ in terms of $\varepsilon$ can be read off by plugging in these scalings into \eqref{eq:scaling-momenta}. On each face $f$ of $\mathbf{P}$, we keep only the terms of $\G$ leading in $\varepsilon$. The result is called the \emph{initial form} and denoted by $\G_f$. The Landau equations on this face are
\be\label{eq:incidence-variety}
\G_f = \partial_{\alpha} \G_f = 0\quad \text{for}\quad \alpha \in (\mathbb{C}^\ast)^m \;\text{and}\; z \in \mathcal{E} \, .
\ee
The system \eqref{eq:LE} is recovered in the special case when $f$ is the full-dimensional face, i.e., the entire polytope. Mathematically, the faces of $\mathbf{P}$ correspond to torus orbits in a \emph{toric compactification} of $(\mathbb{C}^*)^m$. The \emph{incidence variety} $Y_f$ (also called ``pinch surface'') is the variety defined by~\eqref{eq:incidence-variety}. 

Some of the faces can be matched onto the textbook classification, though others are new. For example, the facet with weight $w = (-1,-1,\ldots,-1)$ computes the leading first-type singularity. Replacing $-1$'s with $0$'s and $1$'s corresponds to various subleading singularities, see \cite[Sec.~3.5]{PrincipalLandauDeterminants}. All other faces go beyond the standard classification (note that they also appear in the geometric approach to the expansion by regions \cite{Jantzen:2012mw}). By definition, solutions with all $\alpha_i \geq 0$ are on the physical sheet of the kinematic space \cite{Hannesdottir:2022bmo}; we are interested in singularities on any sheet where $\alpha_i$'s are generically complex. We can further categorize the solutions depending on which value of the external kinematics they occur as follows.

\paragraph*{\bf (a) Any value of the external kinematics, $z \in \mathcal{E}$.} These are the UV/IR divergences, which below are called \emph{dominant components} (also known as ``permanent pinches''). After identifying from which part of the integrand they arise, we want to strip them away since they are instead taken care of by dimensional regularization.

\paragraph*{\bf (b) Special values of the external kinematics, $z \in \{ \Delta  = 0 \} \subset \mathcal{E}$.} These are the kinematic singularities we want to identify. Since analytic functions can only have codimension-$1$ singularities in $\mathcal{E}$, i.e., the ones that can be written as $\Delta=0$, we focus on them here.

A given face can have none, either, or both types of solutions which need to be distinguished. We now move on to studying the geometry of this problem.

\section{Geometry of singularities}

\paragraph{\bf Euler discriminants.} Intuitively, Feynman integrals \eqref{eq:UF-integral} may develop singularities whenever the surface $\{\G(\alpha;z) = 0 \}$ becomes \emph{more singular} than for a generic $z \in {\cal E}$. As a way of probing the topology of this surface, we compute the (signed) Euler characteristic:
\be
\chi_z := \big|\chi\big( (\mathbb{C}^\ast)^m \setminus \{ \G(\alpha;z) = 0 \} \big)\big|\, .
\ee
While non-trivial, one can prove that there exists a generic value $\chi^\ast$, such that $\chi_z = \chi^\ast$ for almost all $z \in \mathcal{E}$ and $\chi_z < \chi^\ast$ otherwise \cite[Thm.~3.1]{PrincipalLandauDeterminants}. This leads us to define the \emph{Euler discriminant} variety $\nabla_{\chi}(\mathcal{E})$ as the locus of such special kinematic points:
\be\label{eq:Euler-discriminant}
\nabla_{\chi}(\mathcal{E}) = \{ \text{all } z \in \mathcal{E} \text{ for which } \chi_z < \chi^\ast \} \subset \mathcal{E}\, .
\ee
If the coefficients of all monomials in $\G$ were independent of each other, $\nabla_{\chi}(\mathcal{E})$ coincides with the singularity locus of generic Gelfand--Kapranov--Zelevinsky (GKZ) hypergeometric functions, which is known as the \emph{principal $A$-determinant} \cite{gelfand1990generalized}. We propose that this relationship extends to Feynman integrals: that their singularities are contained in $\nabla_{\chi}(\mathcal{E})$.

This claim is physically intuitive: $\chi_z$ computes the number of master integrals in the appropriate sense (see \cite{Agostini:2022cgv,Matsubara-Heo:2023ylc} for reviews). The above proposal equivalently says that the system of first-order differential equations satisfied by \eqref{eq:family} drops rank when evaluated on a singularity.

There are two obstacles with applying the above tools to Feynman integrals. Firstly, beyond the simplest of cases, they are not generic in the GKZ sense: for example, the $\G$ polynomial of the diagram in \eqref{eq:contraction} has $64$ monomials, but depends only on $7$ kinematic invariants. Hence, one cannot directly apply GKZ technology such as principal $A$-determinants to Feynman integrals (see \cite{Klausen:2021yrt,Dlapa:2023cvx} for previous attempts). Explicit criteria are given in \cite[Sec.~2.2]{PrincipalLandauDeterminants}. Secondly, while the Euler discriminant can be used to determine whether a given candidate point $z$ is singular, it does not give a constructive algorithm for finding $\nabla_{\chi}(\mathcal{E})$. Both of these problems lead us to the definition of the \emph{principal Landau determinant} (PLD) \footnote{The name \emph{Principal Landau determinant} is inspired by the term \emph{principal A-determinant} coined by Gelfand, Kapranov and Zelevinsky (GKZ) in their study of hypergeometric functions, see for instance \cite[Ch.~10]{gelfand2008discriminants}. The principal $A$-determinant in turn owes its name to its efficient determinantal representation. The GKZ framework considers all coefficients of the polynomial involved as independent parameters. When passing from that generic polynomial to the specialized case of the graph polynomial of a Feynman integral, we replace ``$A$-determinant" by ``Landau determinant" in recognition of the early work of Landau on singularities of Feynman integrals \cite{Landau:1959fi}.}.

\paragraph{\bf Principal Landau determinants.}

The PLD aims to systematize Landau analysis based on the combinatorics of the polytope $\mathbf{P}$ introduced above. A complication is that each face $f$ can produce multiple components and one has to distinguish if they belong to (a) or (b).

The \emph{incidence variety} $Y_f$ defined by \eqref{eq:incidence-variety} is a union of subvarieties of $(\mathbb{C}^\ast)^m \times \mathcal{E}$. In general, it can have multiple irreducible components with different dimensions:
\be
Y_f = \bigcup_i\, Y_f^{(i)}\, .
\ee
Note that the dimension of $Y_f^{(i)}$ is in general different from the dimension of $f$.
We are interested in eliminating $\alpha$'s, which is the same as projecting all components of $Y_f$ ``downstairs'' to the kinematic space $\mathcal{E}$. In other words, for every point $(\hat{\alpha};\hat{z})$ contained in $Y_f$, its projection $\pi(\hat{\alpha};\hat{z}) = \hat{z}$ ``forgets'' about $\hat{\alpha}$ and remembers only the coordinates $\hat{z} \in \mathcal{E}$. We illustrate this with a simple example:
\be
\begin{gathered}
\begin{tikzpicture}
    \node (img) {\includegraphics[valign=c,width=0.7\columnwidth]{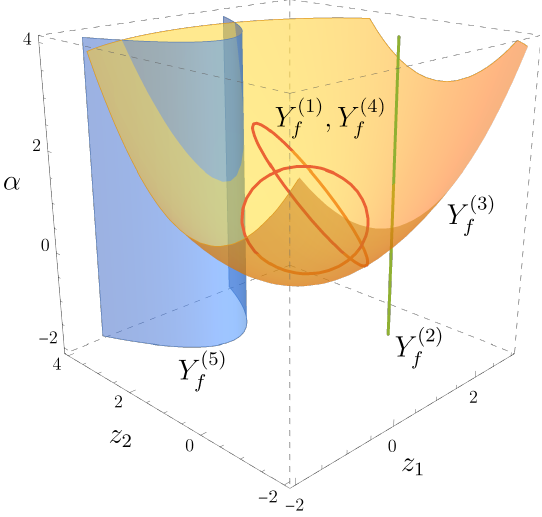}};
    \node [below left, xshift=3.8cm, yshift=0.5cm] at (img) {$\pi$};
    \draw[->] (3.3,1) -- (3.3,-0.5);
\end{tikzpicture}
\end{gathered}
\ee
Here, $\alpha$ and $z = (z_1, z_2)$ are coordinates on the one-dimensional $\mathbb{C}^\ast$ and two-dimensional $\mathcal{E}$ respectively. The incidence variety $Y_f$ has five irreducible components, $Y_f^{(1)}, Y_f^{(2)}, \ldots, Y_f^{(5)}$. Let us analyze them one by one. Below, when we write \emph{fiber dimension} of a component, we mean the dimension of the pre-image $\pi^{-1}(z)$ for a generic point $z$ on $\pi(Y_f^{(i)})$, i.e., it is the number of unconstrained Schwinger parameters remaining after solving \eqref{eq:incidence-variety}.

The \textcolor{orange!80!black}{orange} component $Y_f^{(3)}$ is dominant because it projects down to the whole kinematic space $\mathcal{E}$. It has fiber dimension $0$, i.e., every point downstairs comes from a projection of a unique point upstairs. In physics terms, it corresponds to a UV and/or IR divergence. In other words, the Landau equations \eqref{eq:incidence-variety} have a solution for any $z \in \mathcal{E}$. This, however, does not mean that we can simply disregard the (fictional) face $f$:  there are other irreducible components of type (b).

The \textcolor{Maroon}{red} components $Y_f^{(1)}$ and $Y_f^{(4)}$ both have fiber dimension $0$ and project down to codimension $1$ in $\mathcal{E}$. Next, the \textcolor{RoyalBlue}{blue} component $Y_f^{(5)}$ has fiber dimension $1$ and projects to codimension $1$. We hence keep all three of them in PLD. Finally, the \textcolor{color5}{green} component $Y_f^{(2)}$ has fiber dimension $1$, but projects to codimension $2$. We do not include it in PLD. See \cite[Ex.~5.1]{PrincipalLandauDeterminants} for more details.

Let us call the result of the projection $\pi(Y_f^{(i)})$. For any Feynman diagram $G$, PLD is defined as the union of all those projections that have codimension $1$:
\be
\text{PLD}_G(\mathcal{E}) := \bigcup_{\mathrm{faces }\,f} \bigcup_{\substack{\text{codim-}1 \\ \mathrm{projections }\, i}} \!\!\!\!\! \pi (Y_f^{(i)}) \;\subset \mathcal{E}\, .
\ee
This definition can be used in practice and is implemented in \texttt{PLD.jl} (we refer to \cite[Sec.~3.2]{PrincipalLandauDeterminants} for a formulation in the language of algebraic geometry).

\section{Implementation}

\paragraph{\bf Code and database.} Classic symbolic elimination tools, such as Gr\"obner bases, can be used, but are not efficient enough to handle multi-loop examples. We hence introduce a numerical algorithm for computing PLD, based on homotopy continuation techniques. It performs the above irreducible decomposition and projection to the kinematic space. We refer to \cite[Sec.~5.2]{PrincipalLandauDeterminants} for the algorithmic details. We implemented both symbolic and numerical elimination algorithms in an open-source \texttt{Julia} package \texttt{PLD.jl} available at \cite{MathRepo}. This website also contains documentation and a guided tutorial through the functionality of \texttt{PLD.jl}.
Together with the package, we provide a database of $114$ example diagrams with various graph topologies and mass assignments.

\paragraph{\bf Comparison with \texttt{HyperInt}.} The only competitive tool for finding Landau singularities is the compatibility graph reduction algorithm \texttt{cgReduction} implemented in \texttt{HyperInt} \cite{Panzer:2014caa} and based on the formalism of Pham and Brown in integer dimensions $\D$ \cite{pham2011singularities,Brown:2009ta}. It computes an upper bound on the set of singularities, i.e., finds many components that are not genuine singularities of the corresponding Feynman integral. However, they can be efficiently filtered out by the Euler characteristic criterion \eqref{eq:Euler-discriminant}, see \cite[App.~A]{PrincipalLandauDeterminants} for a practical example. We found diagrams for which this pipeline gave more singularities than those computed numerically by PLD, which means that $\text{PLD}_G(\mathcal{E}) \subsetneq \nabla_{\chi}(\mathcal{E}) \subsetneq \mathtt{cgReduction}$. 
Indeed, it is known that the blow-ups implicitly performed by looking at the Newton polytope might not suffice to detect all singularities of Feynman integrals, see \cite[Sec.~6.4]{Berghoff:2022mqu} and \cite[Ex.~3.9]{PrincipalLandauDeterminants}. In other words, there are scalings of Schwinger parameters that go beyond \eqref{eq:scaling} with genuinely new singularities. These are captured by the Euler discriminant but not by PLD.

On the other hand, in practice, \texttt{cgReduction} terminated only on the simplest $70$ out of $114$ diagrams we considered, so PLD remains the only practical tool for obtaining (a subset of) singularities of more complicated diagrams. At any rate, most of the entries in the database give new predictions for singularities of previously unstudied diagrams.

\section{Example}

As a concrete example, let us consider the following Higgs + jet  production process:
\be
\begin{gathered}\label{eq:diagram}
\begin{tikzpicture}[line width=1,scale=1.3]
	\coordinate (v1) at -(0.2,-0.2);
    \draw (v1) node[above, xshift=2, yshift=2] {\footnotesize{$3$}};
	\coordinate (v7) at -(-0.3,0.5);
    \draw (v7) node[right, xshift=2] {\footnotesize{$2$}};
	\coordinate (v2) at -(0.2,1.2);
    \draw (v2) node[below, xshift=2, yshift=-2] {\footnotesize{$1$}};
	\coordinate (v3) at -(1,1);
    \draw (v3) node[below, xshift=-2, yshift=-2] {\footnotesize{$7$}};
	\coordinate (v4) at -(2,1); \draw (v4) node[below, xshift=2, yshift=-2] {\footnotesize{$5$}};
	\coordinate (v5) at -(2,0); \draw (v5) node[above, xshift=2, yshift=2] {\footnotesize{$4$}};
	\coordinate (v6) at -(1,0); \draw (v6) node[above, xshift=-2, yshift=2] {\footnotesize{$6$}};
	\draw[gluon_small] (v6) -- (v1) -- (v7) -- (v2) -- (v3);
	\draw[quark] (v3) -- (v4) -- (v6);
	\filldraw[white] (1.5,0.5) circle (4pt);
	\draw[quark] (v3) -- (v5);
	\draw[quark] (v5) -- (v6);
	\draw[gluon_small] (v1) -- ++(-135:0.5);
    \draw[<-,Maroon] (0.2-0.2,-0.2) -- node[above left] {\footnotesize{$p_3$}} ++(-135:0.5);
	\draw[gluon_small] (v2) -- ++(135:0.5);
    \draw[<-,Maroon] (0.2-0.2,1.2) -- node[below left] {\footnotesize{$p_1$}} ++(135:0.5);
	\draw[Higgs] (v4) -- ++(45:0.6);
    \draw[<-,Maroon] (2+0.2,1) -- node[below right] {\footnotesize{$p_5$}} ++(45:0.5);
	\draw[gluon_small] (v5) -- ++(-45:0.5);
    \draw[<-,Maroon] (2+0.2,0) -- node[above right] {\footnotesize{$p_4$}} ++(-45:0.5);
	\draw[gluon_small] (v7) -- ++(-180:0.5);
    \draw[<-,Maroon] (-0.3,0.5+0.2) -- node[left] {\footnotesize{$p_2$}\;\;\;\;} ++(-180:0.5);
	\foreach \point in {v1, v2, v3, v4, v5, v6, v7} {
		\fill[white!20!black] (\point) circle [radius=0.07];
	}
\end{tikzpicture}
\end{gathered}
\ee
We take the external Higgs (dashed) to have mass-squared $M^2 = p_5^2$, the quarks (solid) to have mass-squared $m^2$, and all the remaining particles massless. The $m=0$ case has been computed in \cite{Abreu:2021smk}, but $m \neq 0$ remains unknown.
Here, we use \texttt{PLD.jl} to predict its singularity structure.

In this case, the kinematic space ${\cal E}$ is parametrized by
\be
z = (s_{12}, s_{23}, s_{34}, s_{45}, s_{51}, m^2, M^2)\, ,
\ee
where $s_{ij} = (p_i + p_j)^2$ are the Mandelstam invariants.
The diagram is specified by providing the internal \texttt{edges} as a list of pairs of vertices they connect, as in \eqref{eq:diagram}, and similarly \texttt{nodes} listing the vertices to which the external momenta $p_1, p_2, \ldots, p_5$ are attached. Likewise, the \texttt{internal\_} and \texttt{external\_masses} are also assigned in order of appearance. One computes the PLD by running:
\begin{minted}{julia}
edges = [[1,2],[2,3],[3,6],[4,6],
         [5,6],[5,7],[1,7],[4,7]]
nodes = [1,2,3,4,5]

getPLD(edges, nodes,
       internal_masses = [0,0,0,m2,m2,m2,0,m2],
       external_masses = [0,0,0,0,M2])
\end{minted}
The code scans through all faces in increasing order of dimensions (roughly speaking, easier to harder). The generic Euler characteristic is $\chi^\ast = 330$ and the number of faces of $\mathbf{P}$ in dimensions $0,1,\ldots,8$ is
\be
56,\; 294,\; 681,\; 884,\; 699,\; 343,\; 101,\; 16, 1,
\ee
for a total of $3075$ separate systems of equations to consider. Note that a naive analysis based on reduced diagrams would include only $2^8 = 256$ systems, many of which ambiguous as in \eqref{eq:contraction}. We solved the simplest $1270$ ones symbolically and the remaining $1805$ numerically \footnote{The full computation took $\sim 7.2$ hours on two Intel Xeon E5-2695 v4 CPUs with 18 cores each.}, while \texttt{cgReduction} did not terminate.

\texttt{PLD.jl} found the total of $71$ distinct kinematic singularities $\{ \Delta_i = 0\}$ with degrees between $1$ and $12$ in $z$. In general, multiple different faces $f$ contribute to the same singularity. The full list is provided in the $G = \texttt{Hj-npl-pentb}$ entry in our online database. It is a new prediction for the singularities of \eqref{eq:diagram}, though we do not claim that the list is exhaustive.  Here, we give a few examples of $\Delta_i$ for illustration. 

For example, a $2$-dimensional face with weight $(-1, -1, 1, 1, -1, -1, 1, 1)$ contributes a  dominant component together with
\be
\Delta_{1} = M^2\,,\qquad \Delta_{4} = M^2 - 4m^2\, ,
\ee
each coming from $1$-dimensional fibers. They have Euler characteristics $\chi_{\Delta_1} = 244$ and $\chi_{\Delta_4} = 290$, respectively.
Another example is a $3$-dimensional face with weight $-(3, 1, 0, 1, 1, 1, 1, 1)$, giving the degree-$3$ component
\be
\Delta_{15} = 4 m^2 (M^2 s_{23} - s_{45} s_{15}) -s_{15}^2 s_{14} \, ,
\ee
which comes from a $0$-dimensional fiber and has Euler characteristic $\chi_{\Delta_{15}} = 328$. 

The most complicated components we find are $\Delta_{29}$ and $\Delta_{47}$, which both have degree $12$ in $z$. They originate from $0$-dimensional fibers on $5$-dimensional faces with weights  $-(0, 1, 0, 1, 1, 1, 1, 1)$ and $-(1, 0, 1, 1, 1, 1, 0, 1)$ respectively and $\chi_{\Delta_{29}} = \chi_{\Delta_{47}} = 329$. Some terms of $\Delta_{29}$ are
\begin{align}
\Delta_{29} = &116 m^4 M^{10} s_{12}^3 s_{34}^2 - 676 m^2 M^{10} s_{12} s_{34}^4 s_{45} \nn\\
&+ 736 m^4 M^6 s_{12}^3 s_{34} s_{45}^3 - 2656 m^6 M^4 s_{12} s_{34}^2 s_{45}^4 \nn \\
&+m^4 s_{12}^2 s_{34}^4 s_{45}^4 + \text{(329 more terms)}\,.\label{eq:Delta-29}
\end{align}
Similarly, $\Delta_{47}$ is a polynomial with $420$ terms.
Both of them set out our prediction for the most complicated (even) letters of the symbol alphabet in the \eqref{eq:diagram} topology.

We can make a comparison with the $m=0$ case. As emphasized throughout \cite{PrincipalLandauDeterminants}, specializing kinematic parameters and computing PLD in general do not commute, but one can obtain a \emph{subset} of the $m=0$ singularities by substituting $m=0$ in each $\Delta_i$. E.g., \eqref{eq:Delta-29} simplifies drastically~to
\be\label{eq:Delta-29-m0}
\Delta_{29} \big|_{m=0} = M^4 s_{34}^2 (M^4 - M^2 (s_{12} {-} s_{34} {-} s_{45}) + s_{12} s_{45})^4\, .
\ee
Indeed, each factor in this polynomial was already identified as a singularity of the $m=0$ specialization \cite{Abreu:2021smk} and similarly for $\Delta_{47}|_{m=0}$. This can also be verified with the \texttt{EulerDiscriminantQ} function in \texttt{PLD.jl}.
The fact that \eqref{eq:Delta-29} is so much larger than \eqref{eq:Delta-29-m0} suggests that the $m\neq 0$ Feynman integral may have a vastly more complicated analytic form. 

\section{Discussion}

In this letter, we reformulated Landau analysis in a way that can be applied to multi-loop computations in the Standard Model and made it practical by introducing \texttt{PLD.jl}. It was important to realize that singularities can arise from more complicated patterns of shrinking/expanding edges than just reduced diagrams. Previous results relying on this assumption should be revisited. Likewise, our work emphasizes the need to study asymptotics going beyond \eqref{eq:scaling} more carefully.

Even though our discussion focused on Feynman integrals, \texttt{PLD.jl} can be applied to analyzing singularities of other integrals appearing throughout physics. Notable applications include (a) cosmological wavefunctions \cite{Arkani-Hamed:2023kig}, (b) energy correlators in jet physics \cite{Yan:2022cye,Craft:2022kdo}, (c) post-Minkowskian and Newtonian expansions in gravitational-wave physics \cite{Buonanno:2022pgc}, or (d) elliptic and Calabi--Yau geometries arising in perturbative computations \cite{Bourjaily:2022bwx}.

This progress should be viewed as a necessary step in a more ambitious program of using Landau singularities to systematically constrain Feynman integrals relevant to collider and gravitational-wave physics. For example, expanding around incidence varieties would give information about the local nature of the singularity and their discontinuities in the kinematic space, see, e.g., \cite{Hannesdottir:2021kpd,Hannesdottir:2022bmo} for progress on $0$-dimensional fibers without UV/IR subdivergences. We leave these important questions for future work.

\paragraph*{\bf Data Statement.}
The supporting data for this article are openly available from the research data repository MathRepo \cite{MathRepo}.

\paragraph*{\bf Acknowledgements.}
We are grateful to anonymous referees for their insightful comments on a previous version of this letter. We thank Samuel Abreu, Nima Arkani-Hamed, Marko Berghoff, Francis Brown, Hofie Hannesdottir, Johannes Henn, Erik Panzer, Bernd Sturmfels, and Simone Zoia for useful discussions.  C.F. has received funding from the European Union’s Horizon 2020 research and innovation programme under the Marie Sklodowska-Curie grant agreement No 101034255. S.M. gratefully acknowledges funding provided by the Sivian Fund and the Roger Dashen Member Fund at the Institute for Advanced Study.
This material is based upon work supported by the U.S. Department of Energy, Office of Science, Office of High Energy Physics under Award Number DE-SC0009988.

\bibliography{references}
\bibliographystyle{JHEP}

\end{document}